\documentclass[preprintnumbers,preprint,amsmath,amssymb]{revtex4}
\usepackage{amssymb}
\usepackage[colorlinks, linkcolor=blue, citecolor=red, urlcolor=red]{hyperref}
\usepackage{graphicx}
\usepackage{dcolumn}
\usepackage{bm}

\newcommand{\bea}{\begin{eqnarray}}
\newcommand{\eea}{\end{eqnarray}}

\begin{document}

\title{Is the Cosmic Transparency Spatially Homogeneous?}

\author{Jun Chen$^{a,b}$,  Puxun Wu$^{c}$, Hongwei Yu$^{a, c,} \footnote{Corresponding author: hwyu@hunnu.edu.cn}$
  ~and Zhengxiang Li$^{a}$ }

\address{$^a$Institute of Physics and Key Laboratory of Low
Dimensional Quantum Structures and Quantum Control of Ministry of
Education, Hunan Normal University, Changsha, Hunan 410081, China
\\$^b$Department of Physics and Electronic Science, Hunan University of Arts and Science, Changde, Hunan 415000, China
\\$^c$Center for Nonlinear Science and Department of Physics, Ningbo
University,  Ningbo, Zhejiang 315211, China }

\begin{abstract}
 We study the constraints on the cosmic opacity using the latest BAO and Union2 SNIa data in this paper and find that the best fit values seem to indicate that an opaque universe is preferred  in redshift regions $0.20-0.35$, $0.35-0.44$ and $0.60-0.73$, whereas, a transparent universe is favored in redshift regions $0.106-0.20$, $0.44-0.57$and $0.57-0.60$. However, our result is still consistent  with a transparent universe  at the  
1$\sigma$ confidence level,  
 even though 
the best-fit cosmic opacity oscillates between zero and some nonzero values as the redshift varies.
\end{abstract}

\maketitle

\section{INTRODUCTION}
  The present accelerating cosmic expansion is discovered firstly  from the observation of Type Ia supernova (SNIa)~\cite{Riess1998, Perlmutter1999} since it revealed that the supernova are  fainter than  expected from a decelerating expanding universe under the assumption that the universe is transparent.   However, the universe may be opaque due to that there may exist sources for photon attenuation, such as absorption or scattering of gas and plasma in and around galaxies.
Furthermore, if dark matter is an axion or axion-like particle,  photons propagating in extragalatic magnetic fields may oscillate into  axion or axion-like particle. It has been shown that this  axion-photon mixing may account for the dimming of SNIa~\citep{Aguirre, Csaki} without the need of dark energy or modified gravity.

 The photon attenuation  inevitably leads to a violation of the distance duality (DD) relation, also called the Etherington relation~\citep{Etherington}:
 \begin{equation}\label{ddrelation}
 D_{L}=(1+z)^{2}D_{A},
\end{equation}
 which is built on two assumptions: the conservation of photon number and  Lorentz invariance.  Here,  $z$ is the redshift,  $D_L$ and $D_A$ are the luminosity distance and the angular diameter distance, respectively. If one can have both $D_A$ and $D_L$ at the same redshift from observations, the DD relation can be tested  directly~\cite{Uzan, Bernardis, Avgoustidis, Holandaa, Holandab, Nair, Liang, Lima, Meng, Cao, Holandac,Holandad, Goncalves, Holandae, Khedekar}.   From the Union2 SNIa, which provides the luminosity distance, and the galaxy cluster data giving the angular diameter distance, it has been found that the DD relation is consistent with the observations at the $1\sigma$ confidence level when the elliptical galaxy  cluster data~\citep{Filippis} is used. However, the consistency occurs only at the $3\sigma$ confidence level~\citep{Li} for the spherical galaxy cluster data~\citep{Bonamente}.  Thus, a violation of the DD relation is not excluded.  This may  be considered as a result of a breakdown of physics on which the DD relation is based upon~\citep{Csaki, Bassetta, Bassettb}.

Assuming that the photon travels on null geodesics, the cosmic opacity becomes the only source for the breakdown of  the DD relation. Thus,  observational data on $D_L$ and $D_A$  can give a constraint  on the cosmic opacity.
 \citet{More} have used the Sloan Digital Sky Survey  (SDSS)  and the Two Degree Field Galaxy Redshift Survey (2dFGRS) BAO  data measured at redshifts $z=0.35$ and $0.20$~\citep{Percival} and SNIa data from \citep{Davis} to constrain the opacity, and found that a transparent universe is favored and an opacity $\Delta \tau<0.13$ at the $95\%$ confidence level for the redshift range between $0.2$ and $0.35$.  Later, \citet{Avgoustidis}  studied the   possible deviation from transparency  by using the cosmic expansion history $H(z)$ data~\citep{Stern} and the Union SNIa data~\citep{Kowalski}, and  found $\Delta \tau<0.012$ at the $95\%$ confidence level between redshift $0.2$ and $0.35$, which is a factor of $2$ stronger than what was obtained in \cite{More}.

Recently, the 6-degree Field Galaxy Survey (6dFGS) has reported a BAO detection in the low-redshift universe $z=0.106$~\citep{Beutler}.  The WiggleZ Dark Energy Survey has released the baryon acoustic peak at redshifts z = 0.44, 0.6 and 0.73~\citep{Blake}, and the Baryonic Oscillation Spectroscopic Survey (BOSS) has given  a data point at $z=0.57$~\citep{Sanchez}. Combining the WiggleZ dark energy Survey with 6dFGS, SDSS and BOSS, we  now have seven BAO data points. Except for the data from BOSS, the other six data points  have been  shown in Tab. (3) in~\cite{Blake}.  Furthermore, the SDSS BAO survey has released the latest  results~\citep{Percivalb}. Therefore, it is of interest to re-examine the cosmic opacity using these latest BAO data
 and this is what we are going to do in this paper.  The behavior of cosmic opacity in different redshift regions will be discussed in detail.

\section{THE COSMIC TRANSPARENCY}
It follows from Eq.~(\ref{ddrelation}) that the transparency of the universe requires
\begin{eqnarray}\label{dadl}
\frac{(1+z_{2})^{2}}{(1+z_{1})^{2}}\frac{D_{A}(z_{2})}{D_{A}(z_{1})}=\frac{D_{L}(z_{2})}{D_{L}(z_{1})}\;,
\end{eqnarray}
which  is independent of cosmological models and  so far, it has been applied to all analysis of the cosmological observations without any doubt. Its validity can however be tested observationally. In this regard, let us note that since BAO
provides a standard ruler for direct measurement of the
cosmic expansion history, therefore, we can obtain the angular
diameter distance from  BAO observation which is independent of photon attenuation. For BAO data, the acoustic
parameter $A(z)$ introduced by \citet{Eisenstein}
\begin{equation}\label{Az}
A(z)=\frac{100 D_V(z)\sqrt{\Omega_mh^2}}{cz}\;,
\end{equation}
is used usually,  where $\Omega_m$ is the matter density parameter, $h=H_0/100$, and the hybrid
distance $D_{V}$ relates with the angular diameter distance $D_{A}$ through
\begin{equation}\label{L1}
 D_{V}=\bigg(\frac{cz(1+z)^{2}D_{A}^{2}}{H(z)}\bigg)^{\frac{1}{3}}.
\end{equation}
Here $H(z)$ is the Hubble expansion rate at redshift $z$.
To test Eq.~(\ref{dadl}),  we use, for convenience, the ratio of $A$ instead of $D_L$.
From  Tab.~(3) in \cite{Blake} and the data from BOSS~\cite{Sanchez}, we get the observed ratios of
$A$, which are given in Tab.~(\ref{Taba}).

\begin{table}[!h]\centering
\begin{tabular}{|c|c|c|}
\hline ~ $\frac{A(0.2)}{A(0.106)}$
~~&~~$\frac{A(0.35)}{A(0.2)}$~~&~~$\frac{A(0.44)}{A(0.35)}$~~\\
\hline
$0.928\pm 0.058$ &~~ $0.992\pm 0.046$~~&~~$0.979\pm 0.077$~~\\
\hline ~ $\frac{A(0.57)}{A(0.44)}$ ~~&~~$\frac{A(0.6)}{A(0.57)}$ ~~&~~$\frac{A(0.73)}{A(0.6)}$ ~~\\
\hline
$0.937\pm 0.073$ ~~&~~$0.995\pm 0.055$~~&~~$0.959\pm 0.064$~~\\
\hline
\end{tabular}
\tabcolsep 0pt \caption{\label{Taba} The ratio of the acoustic
parameter $A(z)$ obtained from Refs.~\cite{Blake, Sanchez}.}
\vspace*{5pt}
\end{table}

SNIa has been found to be standard candles which can be used to
make an independent direct measurement of the expansion
history~\citep{Riess1998, Perlmutter1999}. Its luminosity distance
can be determined by measuring energy per unit time per unit area received at a telescope. Thus, we can determine the right-hand side of Eq.~(\ref{dadl})  from  SNIa directly.
In this paper,  the Union2 SNIa sample, which contains 557 data points, is used~\citep{Amanullah}.
In order to obtain the luminosity distance at the corresponding redshift $z$ of BAO data, we bin all SNIa data in the redshift range
$[z-0.05,z+0.05]$ and  the binned $D_L$ at redshift
$z$ is
\begin{equation}
 D_{L}^{bin}=\frac{\sum D_{L_{i}}/(\sigma^{2}_{D_{L_{i}}}+
 \sigma_{S_i}^2)}{\sum 1/(\sigma^{2}_{D_{L_{i}}}+
 \sigma_{S_i}^2)},
\end{equation}
with  $\sigma^{2}_{D^{bin}_{L}}$ being
\begin{equation}
 \sigma^{2}_{D^{bin}_{L}}=\frac{1}{\sum
 1/(\sigma^{2}_{D_{L_{i}}}+
 \sigma_{S_i}^2)}.
\end{equation}
Here, $\sigma_{D_{L_{i}}}$ is  the uncertainty of the individual
distance and $\sigma_{S_i}$ is the corresponding systematic error. For the Union2 SNIa sample,  systematic errors have been compiled in Tab.~(7)
of \cite{Amanullah}.

If the universe is opaque, that is, there are some sources for photon attenuation, the observed luminosity  distance derived from SNIa will be modified and it will be larger than the true one. Let $\tau(z)$ denotes the opacity between an observer at redshift $z=0$
and a source at $z$. The flux received from this source would be reduced
by a factor $e^{-\tau(z)}$. The relation between the observed luminosity distance
and the true one becomes~\citep{Chen}:
\begin{equation}
D_{L_{true}}^{2}=D_{L_{obs}}^{2}e^{-\tau(z)}.
\end{equation}
Here, $D_{L_{obs}}$ is obtained from
SNIa data. Since BAO observation is not affected by the photon attenuation,  $D_{L_{true}}$ can be derived  from  BAO data.  If the universe is transparent, $\tau(z)$ is
zero.

Because SNIa observation only releases the distance modulus data, which relates to the luminosity distance through
\begin{equation}
\mu=5\log D_{L}+25,
\end{equation}
the observed distance modulus also differs from the true one
\begin{equation}
\mu_{obs}(z)=\mu_{true}(z)+(2.5\log e)\tau(z).
\end{equation}
Thus, the distance modulus difference between two redshifts $z_{1}$ and
$z_{1}$ is
\begin{equation}
\Delta\mu_{obs}=\mu_{obs}(z_{2})-\mu_{obs}(z_{1}),
\end{equation}
and then  one can obtain
\begin{equation}\label{dm}
\Delta\mu_{obs}=5 \log \frac{D_{L_{true}}(z_2)}{D_{L_{true}}(z_1)}+2.5 \Delta\tau\log e ,
\end{equation}
where $\Delta\tau=\tau(z_2)-\tau(z_1)$.  From the Union2 SNIa data and the binning method, we find  $\Delta\mu_{obs}$ at the redshift differences of BAO data and show them in Tabs.~(\ref{Tabdm}) and (\ref{Tabdm2}), which correspond to the case without and with systematic errors, respectively.

\begin{table}[!h]\centering
\begin{tabular}{|c|c|c|}
\hline ~ $\mu_{obs}(0.2)-\mu_{obs}(0.106)$
~~&~~$\mu_{obs}(0.35)-\mu_{obs}(0.2)$
~~&~~$\mu_{obs}(0.44)-\mu_{obs}(0.35)$ ~~\\
\hline
$1.332\pm 0.094$ &~~ $1.439\pm 0.101$~~&~~$0.606\pm 0.116$~~\\
\hline ~ $\mu_{obs}(0.57)-\mu_{obs}(0.44)$
~~&~~$\mu_{obs}(0.6)-\mu_{obs}(0.57)$
~~&~~$\mu_{obs}(0.73)-\mu_{obs}(0.6)$ ~~\\
\hline
$0.533\pm 0.160$ &~~ $0.079\pm 0.188$~~&~~$0.598\pm 0.188$~~\\
\hline
\end{tabular}
\tabcolsep 0pt \caption{\label{Tabdm} The SNIa distance modulus
difference.} \vspace*{5pt}
\end{table}

\begin{table}[!h]\centering
\begin{tabular}{|c|c|c|}
\hline ~ $\mu_{obs}(0.2)-\mu_{obs}(0.106)(sys)$
~~&~~$\mu_{obs}(0.35)-\mu_{obs}(0.2)(sys)$
~~&~~$\mu_{obs}(0.44)-\mu_{obs}(0.35)(sys)$ ~~\\
\hline
$1.332\pm 0.120$ &~~ $1.430\pm 0.123$~~&~~$0.621\pm 0.137$~~\\
\hline ~ $\mu_{obs}(0.57)-\mu_{obs}(0.44)(sys)$
~~&~~$\mu_{obs}(0.6)-\mu_{obs}(0.57)(sys)$
~~&~~$\mu_{obs}(0.73)-\mu_{obs}(0.6)(sys)$ ~~\\
\hline
$0.533\pm 0.195$ &~~ $0.082\pm 0.232$~~&~~$0.601\pm 0.236$~~\\
\hline
\end{tabular}
\tabcolsep 0pt \caption{\label{Tabdm2} The SNIa  distance modulus
difference with systematic errors included.} \vspace*{5pt}
\end{table}

Combining Eqs.~(\ref{Az}, \ref{L1}, \ref{dm}) and using the fact that $D_{L_{true}}$ can be deduced from  BAO data,
we have
\begin{equation}
\Delta\mu_{obs}=\frac{5}{2}\bigg(\frac{\Delta\tau}{\ln(10)}+3\log\frac{A(z_{2})}{A(z_{1})}
-\log\frac{z_{1}^2(1+z_{1})^{2}H(z_{1})}{z_{2}^2(1+z_{2})^{2}H(z_{2})}\bigg).
\end{equation}
The last term on the right hand side of the above equation shows that a cosmological  model must be assumed  to
find the transparency of our universe.  Here,  we consider the
$\Lambda$CDM model,
$E(z)=\frac{H(z)}{H_0}=\sqrt{\Omega_{m}(1+z)^{3}+\Omega_{\Lambda}+\Omega_{k}(1+z)^{2}}$ with $\Omega_{k}=1-\Omega_{\Lambda}-\Omega_{m}$. Following Ref.~\cite{More}, we marginalize over $\Omega_m$ and $\Omega_\Lambda$ with
$\Omega_{\Lambda}\in[0,1]$ and $\Omega_{M}\in[0,1]$, and  calculate the posterior
probabilities of $\Delta\tau$ by using the Bayesian approach
\begin{equation}
P(\Delta\tau|S,B)=\int_{\Omega_{\Lambda}}\int_{\Omega_{M}}P(\Omega_{\Lambda},\Omega_{M}|B)
P(\Delta\tau,\Omega_{\Lambda},\Omega_{M}|S)d\Omega_{\Lambda}d\Omega_{M},
\end{equation}
where  \begin{equation}
P(\Omega_{\Lambda},\Omega_{M}|B)=\frac{\exp(-\frac{\chi^{2}_{B}}{2})}
{\int_{\Omega_{\Lambda}}d\Omega_{\Lambda}\int_{\Omega_{M}}d\Omega_{M}
\exp(-\frac{\chi^{2}_{B}}{2})},
\end{equation}
and
\begin{equation}
P(\Delta\tau,\Omega_{\Lambda},\Omega_{M}|S)=\frac{\exp(-\frac{\chi^{2}_{S}}{2})}
{\int_{\Omega_{\Lambda}}d\Omega_{\Lambda}\int_{\Omega_{M}}d\Omega_{M}
\int_{0}^{0.5}d\Delta\tau\exp(-\frac{\chi^{2}_{S}}{2})}
\end{equation}
are the posterior
probabilities of the set of model parameters given by BAO and SNIa
data.
 Assuming that the uncertainties on BAO and SNIa are Gaussian,
we have
\begin{equation}
\chi^{2}_{B}=\frac{1}{\sigma_{obs}^{2}}\bigg(\frac{A(z_{2})}{A(z_{1})}-\frac{A_{obs}(z_{2})}{A_{obs}(z_{1})}\bigg)^{2},
\end{equation}
\begin{equation}
\chi^{2}_{S}=(\Delta\mu_{true}-\Delta\mu_{obs})^{2}/\sigma_{obs}^{2}.
\end{equation}

\begin{figure}[h!]
   \centering
       \includegraphics[width=0.4\linewidth]{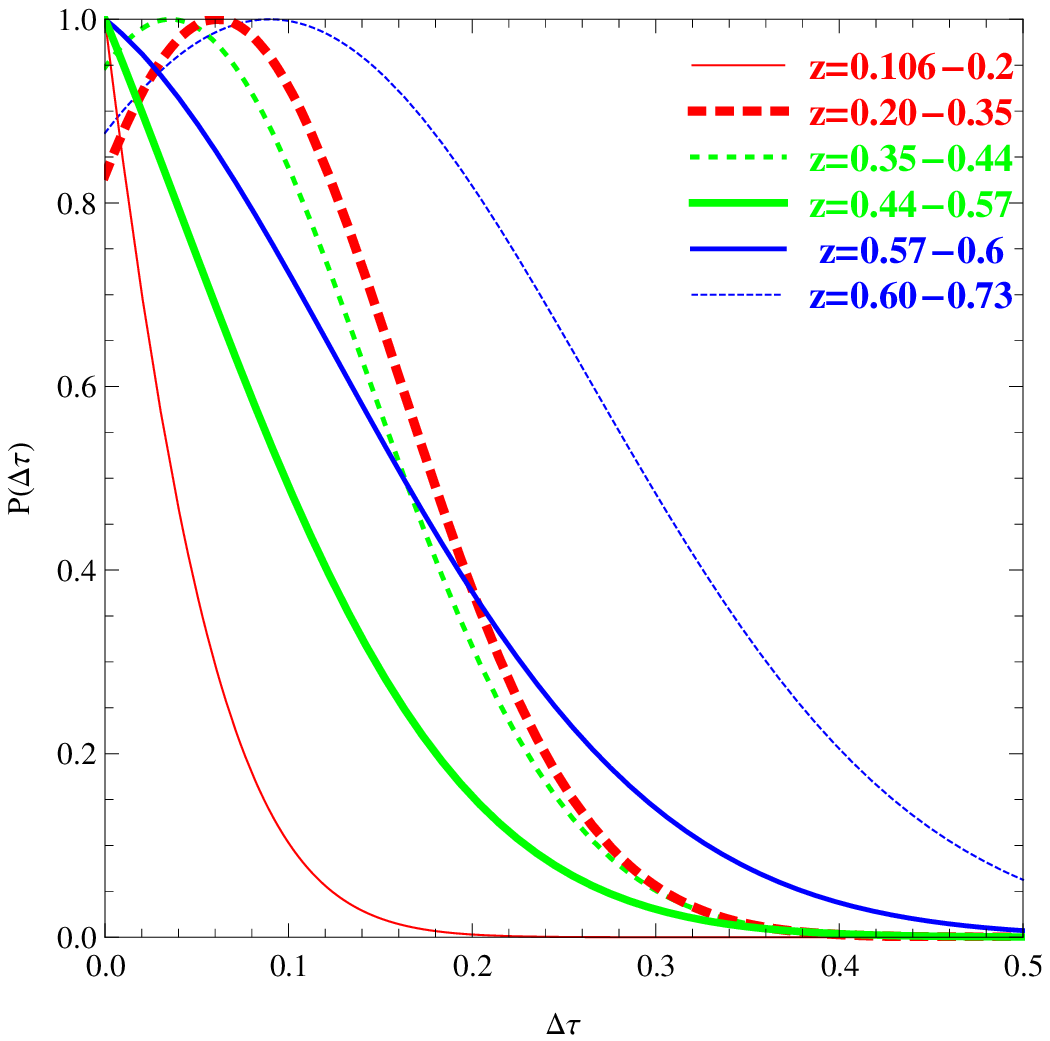}  \includegraphics[width=0.4\linewidth]{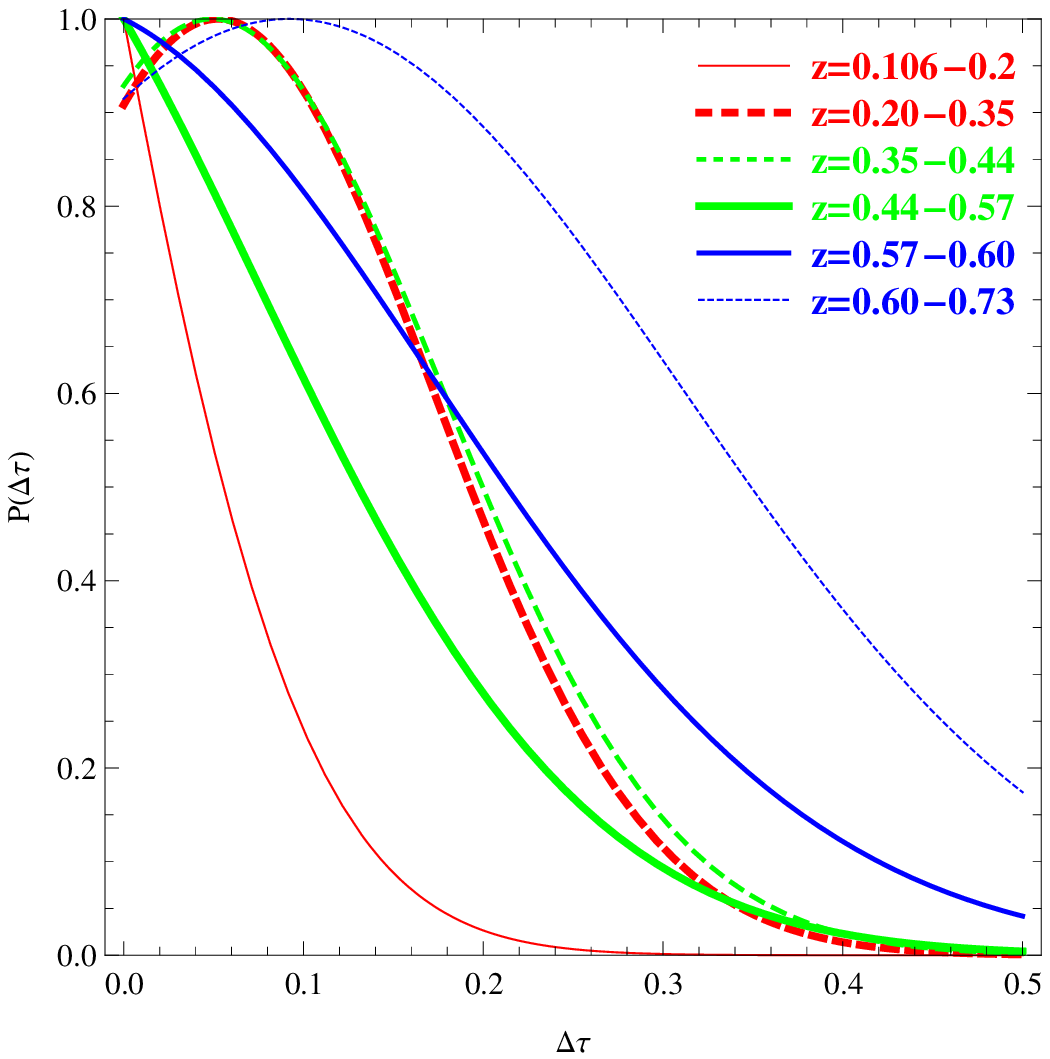}
   \caption{\label{Figpp} The posterior probabilities of $\Delta\tau$. The left and right panels show the results without and with systematic errors, respectively}
\end{figure}

 The constraints can then be obtained and results are shown in  Fig.~(\ref{Figpp}) and Tabs.~(\ref{Tabdtao}, \ref{Tabdtao2}).   The posterior distributions show that the universe is transparent between redshift regions
$0.106-0.2$, $0.44-0.57$ and $0.57-0.6$,  while it seems to be
opaque at  redshift regions $0.2-0.35$, $0.35-0.44$ and $0.6-0.73$ since the best fit values of $\Delta \tau $ are $0.061$, $0.036$ and $0.090$ ($0.052$, $0.049$ and $0.092$ when systematic errors are considered)  at
these redshift  regions, respectively.  However, at the 1$\sigma$ confidence level, $\Delta\tau=0$  is still allowed. Thus, our result is consistent with a transparent universe at the 1$\sigma$ confidence level no matter  whether systematic errors are included or not,   although
  the cosmic opacity seems to show different properties at different redshift regions.     
   
In addition, we find that our result at redshift region $0.2-0.35$ is clearly different from what was obtained in \cite{More} where  a transparent universe is favored and at the $95\%$ confidence level $\Delta\tau<0.13$.
This difference may come from the fact that we use the latest SDSS DR7 BAO data while the SDSS DR5 is considered in \cite{More}. Our result also differs from what was obtained
 using the Hubble data in \cite{Avgoustidis} where $\Delta\tau<0.012$ at the $95\%$ confidence level at redshift region $0.2-0.35$.

 \begin{table}[!h]\centering
\begin{tabular}{|c|c|c|c|c|}
\hline ~ $ $~~&~~best fit value~~&~~$1\sigma$~~&~~$2\sigma$~~&~~$3\sigma$~~\\
\hline ~ $\Delta\tau_{0.106-0.2}$~~&~~0~~&~~$0.043$~~&~~$0.108$~~&~~$0.181$~~\\
\hline ~ $\Delta\tau_{0.20-0.35}$~~&~~ $0.061$ ~~&~~$0.132$~~&~~$0.235$~~&~~$0.342$~~\\
\hline ~ $\Delta\tau_{0.35-0.44}$~~&~~ $0.036$ ~~&~~$0.121$~~&~~$0.233$~~&~~$0.348$~~\\
\hline ~ $\Delta\tau_{0.44-0.57}$~~&~~ $0$ ~~&~~$0.101$~~&~~$0.225$~~&~~$0.362$~~\\
\hline ~ $\Delta\tau_{0.57-0.60}$~~&~~ $0$ ~~&~~$0.152$~~&~~$0.306$~~&~~$0.444$~~\\
\hline ~ $\Delta\tau_{0.60-0.73}$~~&~~ $0.090$ ~~&~~$0.214$~~&~~$0.386$~~&~~$0.479$~~\\
 \hline
\end{tabular}
\tabcolsep 0pt \caption{\label{Tabdtao} The obtained $\Delta\tau$ in different redshift regions.}
\vspace*{5pt}
\end{table}

\begin{table}[!h]\centering
\begin{tabular}{|c|c|c|c|c|}
\hline ~ $ $~~&~~best fit values~~&~~$1\sigma$~~&~~$2\sigma$~~&~~$3\sigma$~~\\
\hline ~ $\Delta\tau_{0.106-0.2} (sys)$~~&~~0~~&~~$0.058$~~&~~$0.153$~~&~~$0.247$~~\\
\hline ~ $\Delta\tau_{0.20-0.35}(sys)$~~&~~ $0.052$ ~~&~~$0.139$~~&~~$0.268$~~&~~$0.390$~~\\
\hline ~ $\Delta\tau_{0.35-0.44}(sys)$~~&~~ $0.049$ ~~&~~$0.146$~~&~~$0.282$~~&~~$0.411$~~\\
\hline ~ $\Delta\tau_{0.44-0.57}(sys)$~~&~~ $0$ ~~&~~$0.129$~~&~~$0.286$~~&~~$0.444$~~\\
\hline ~ $\Delta\tau_{0.57-0.60}(sys)$~~&~~ $0$ ~~&~~$0.184$~~&~~$0.371$~~&~~$0.475$~~\\
\hline ~ $\Delta\tau_{0.60-0.73}(sys)$~~&~~ $0.092$ ~~&~~$0.245$~~&~~$0.423$~~&~~$0.492$~~\\
\hline
\end{tabular}
\tabcolsep 0pt \caption{\label{Tabdtao2}  The obtained $\Delta\tau$ in different redshift regions with systematic errors included in SNIa.}
\vspace*{5pt}
\end{table}

\section{CONCLUSION}
An opaque universe is an  interesting possibility  since it is capable of accounting for the SNIa dimming with no need of an accelerated cosmic expansion. In this paper, we discuss the constraints on the cosmic opacity from  the latest BAO data, released from 6dFGS, SDSS, BOSS and WiggleZ survey,  and  the Union2 SNIa data.  In our discussion, the effect of systematic errors in the Union2 SNIa is considered. The best fit values show that, whether systematic errors are included or not,  the data between the redshift regions $0.106-0.20$, $0.44-0.57$ and $0.57-0.60$ favor a transparent universe,
whereas, when the data between the redshift regions $0.20-0.35$, $0.35-0.44$ and $0.60-0.73$  are used,  an opaque universe is preferred. However, at the $68.3\%$ confidence level, $\Delta\tau=0$ is still allowed by observations.  Our result at the redshift region $0.20-0.35$ is different from what was obtained in \cite{More}  where the SDSS DR5 BAO data  are used and a transparent universe is found. This difference may come from that we use the latest SDSS DR7 BAO data. It also differs from the conclusion drawn from the Hubble  data  between redshift region $0.20-0.35$~\citep{Avgoustidis}.  Although our result shows that the best-fit cosmic opacity oscillates between zero and some nonzero values as the redshift varies,   a transparent universe is  consistent with observations at the 1$\sigma$ 
confidence level.

 \acknowledgments  This work was supported by
the National Natural Science Foundation of China under Grants Nos.
10935013, 11175093,  11222545  and 11075083, Zhejiang Provincial Natural Science
Foundation of China under Grants Nos. Z6100077 and R6110518, the
FANEDD under Grant No. 200922, the National Basic Research Program
of China under Grant No. 2010CB832803, the NCET under Grant No.
09-0144,  the PCSIRT under Grant No. IRT0964, the Hunan Provincial
Natural Science Foundation of China under Grant No. 11JJ7001, and
the Program for the Key Discipline in Hunan Province.

\begin{thebibliography}{}
\bibitem[Perlmutter et al. (1999)]{Perlmutter1999} Perlmutter, S., Aldering, G., Goldhaber, G.,  et al., ApJ 517 (1999) 565
\bibitem[Riess et al. (1998)]{Riess1998} Riess, A. G., Filippenko, A. V., Challis, P., et al., AJ 116 (1998) 1009
\bibitem[Csaki, Kaloper \& Terning (2002)]{Csaki}Csaki, C., Kaloper, N., \& Terning, J.,  Phys. Rev. Lett. 88 (2002) 161302
\bibitem[Aguirre (1999)]{Aguirre}Aguirre, A.,  ApJ 525 (1999) 583
\bibitem[Etherington (1933)]{Etherington}Etherington, J. M. H.,  Phil. Mag. 15  (1933) 761
\bibitem[Avgoustidis et al. (2010)]{Avgoustidis}Avgoustidis, A., Burrage, C., Redondo, J., Verde, L., \& Jimenez, R.,  JCAP 1010 (2010) 024
\bibitem[Cao \& Liang (2011)]{Cao} Cao, S. \& Liang, N., arXiv: 1104.4942
\bibitem[De Bernardis et al. (2006)]{Bernardis}De Bernardis, F., Giusarma, E., \& Melchiorri, A.,  Int. J. Mod. Phys. D 15 (2006) 759
\bibitem[Goncalves et al. (2011)]{Goncalves}Goncalves, R. S.,  Holanda,  R. F. L.,  \& Alcaniz, J. S., MNRAS 420 (2011) L43
\bibitem[Holanda et al. (2010a)]{Holandaa}Holanda, R. F., Lima, J. A. S., \& Ribeiro, M. B.,  arXiv: 1003.5906.
\bibitem[Holanda et al. (2010b)]{Holandab}Holanda, R. F., Lima, J. A. S., Ribeiro, \& M. B.,  ApJ 722 (2010) L233
\bibitem[Holanda et al. (2012a)]{Holandac}Holanda, R. F.,  Goncalves, R. S.,  \&  Alcaniz, J. S.,  arXiv: 1201.2378
\bibitem[Holanda (2012)]{Holandad}Holanda, R. F.,   arXiv: 1202.2309
\bibitem[Holanda et al. (2012b)]{Holandae}Holanda, R. F., Lima, J. A. S., \& Ribeiro, M. B.,  A\&A 538 (2012) A131
\bibitem[Khedekar \& Chakraborti (2011)]{Khedekar}Khedekar, S. \& Chakraborti, S.,  Phys. Rev. Lett. 106 (2011) 221301
\bibitem[Liang et al. (2011)]{Liang}Liang, N.,  Cao, S.,  Liao, K., \& Zhu, Z. H.,  arXiv: 1104.2497
\bibitem[Lima, Cunha \& Zanchin (2011)]{Lima}Lima, J. A. S., Cunha, J. V., \& Zanchin,  V. T.,  arXiv: 1110.5065
\bibitem[Meng et al. (2011)]{Meng} Meng, X.,  Zhang, T., Zhan, H.,  \&  Wang, X.,  ApJ 745 (2012) 98
\bibitem[Nair et al. (2011)]{Nair}Nair, R., Jhingan, S.,  \& Jain, D.,   JCAP 05 (2011) 023
\bibitem[Uzan et al. (2004)]{Uzan}Uzan, J. P., Aghanim, \& N., Mellier, Y.,  Phys. Rev. D 70 (2004) 083533
\bibitem[De Filippis et al. (2005)]{Filippis}De Filippis, E., Sereno,  M., Bautz, W., \& Longo, G.,  ApJ 625 (2005) 108
\bibitem[Li, Wu \& Yu (2011)]{Li} Li, Z., Wu, P., \& Yu, H.,  ApJ 729 (2011) L14
\bibitem[Bonamente et al. (2006)]{Bonamente}Bonamente, M., et al., ApJ 647 (2006) 25
\bibitem[Bassett \& Kunz (2004a)]{Bassetta}Bassett, B. A. \& Kunz, M.,  ApJ 607 (2004) 661
\bibitem[Bassett \& Kunz (2004b)]{Bassettb}Bassett, B. A. \& Kunz, M.,  Phys. Rev. D 69 (2004) 101305
\bibitem[More et al. (2009)]{More} More, S., Bovy, J., \& Hogg, D. W.,  ApJ 696 (2009) 1727
\bibitem[Percival et al. (2007)]{Percival} Percival, W. J., et al., MNRAS 381 (2007) 1053
\bibitem[Davis et al. (2007)]{Davis} Davis, T. M., et al., ApJ 666 (2007) 716
\bibitem[Stern et al. (2010)]{Stern}Stern, D., Jimenez, R., Verde, L., Kamionkowski, M., \& Stanford, S. A.,  JCAP 1002 (2010) 008
\bibitem[Kowalski et al. (2008)]{Kowalski} Kowalski et al., ApJ 686 (2008) 749
\bibitem[Beutler et al. (2011)]{Beutler}Beutler F., et al.,  MNRAS 416 (2011) 3017
\bibitem[Blake et al. (2011)]{Blake} Blake, C., et al.  MNRAS 418 (2011) 1707
\bibitem[Sanchez et al. (2012)]{Sanchez} Sanchez, A. G., et al. arXiv: 1203.6616
\bibitem[Percival et al. (2010)]{Percivalb} Percival, W. J., et al., MNRAS 401 (2010) 2148
\bibitem[Eisenstein et al. (2005)]{Eisenstein}Eisenstein D.J., et al.,  ApJ 633 (2005) 560
\bibitem[Amanullah et al. (2010)]{Amanullah}Amanullah, R., et al., ApJ 716 (2010) 712
\bibitem[Chen \& Kantowski (2009)]{Chen}Chen, B. \&  Kantowski, R.,  Phys. Rev. D 79 (2009) 104007
\end{thebibliography}
\end{document}